# ezcox: An R/CRAN Package for Cox Model Batch Processing and Visualization

Shixiang Wang[1,2], Xue-Song Liu[2], Jianfeng Li[3,4] and Qi Zhao[1,*]

Affiliations of authors:

[1] Bioinformatics Platform, Department of Experimental Research, Sun Yat-sen University Cancer Center, State Key Laboratory of Oncology in South China, Collaborative Innovation Center for Cancer Medicine, Sun Yat-sen University, Guangzhou 510060, China;

[2] School of Life Science and Technology, ShanghaiTech University, Shanghai 201203, China;

[3] State Key Laboratory of Medical Genomics, Shanghai Institute of Hematology, National Research Center for Translational Medicine, Rui-Jin Hospital, Shanghai Jiao Tong University, School of Medicine, Shanghai 200025, China;

[4] School of Life Sciences and Biotechnology, Shanghai Jiao Tong University, Shanghai 200240, China.

*Correspondence: Qi Zhao. Email: zhaoqi@sysucc.org.cn.

**Abstract**

Cox analysis is a common clinical data analysis technique to link valuable variables to clinical outcomes including dead and relapse. In the omics era, Cox model batch processing is a basic strategy for screening clinically relevant variables, biomarker discovery and gene signature identification. However, all such analyses have been implemented with homebrew code in research community, thus lack of transparency and reproducibility. Here, we present *ezcox*, the first R/CRAN package for Cox model batch processing and visualization. *ezcox* is an open source R package under GPL-3 license and it is free available at https://github.com/ShixiangWang/ezcox and https://cran.r-project.org/package=ezcox.

**Introduction**

Large cancer research projects (e.g. TCGA[1], CCLE[2], ICGC[3] and PCAWG[3]), curated cancer omics databases (e.g., UCSC Xena[4] and cBioPortal[5]) and individual cancer research groups have generated huge amounts of cancer omics data for enabling unprecedented research opportunities in silico. Cox model analysis[6] is a common clinical data analysis technique to link valuable variables to clinical outcomes including dead and relapse. In the omics era, Cox model batch processing is a basic strategy for screening clinically relevant variables, biomarker discovery and gene signature identification[7]–[9]. However, all such analyses have been implemented with homebrew code in research community, thus lack of transparency and reproducibility. Furthermore, flexible and publication-ready plot for result data from batch processing is generally lacking. *ezcox* is born to address these issues for open science, especially in cancer research areas.

**Methods and results**

*ezcox* has been developed with R>=3.5 following a modular and robust design of R package. Continuous integration testing with CRAN R package check standards is done automatically after each code commit to help test functionality and detect program bugs in a timely manner. Instructions on how to install, use *ezcox* are presented in the public GitHub repository (https://github.com/ShixiangWang/ezcox).

The architecture of *ezcox* can be classified into two parts (Fig. 1). The first part contains basic batch processing steps: 1) construct multiple univariable/multivariable Cox models in batch with sequential or parallel execution, then store the models along with key parameters (e.g., hazard ratio and its 95% confidence interval, p value; Fig. 2A) in tidy way; 2) filter result and retrieve corresponding models, e.g., only keep models with statistically significant result; 3) show specified models with forest plot (Fig. 2B-D).

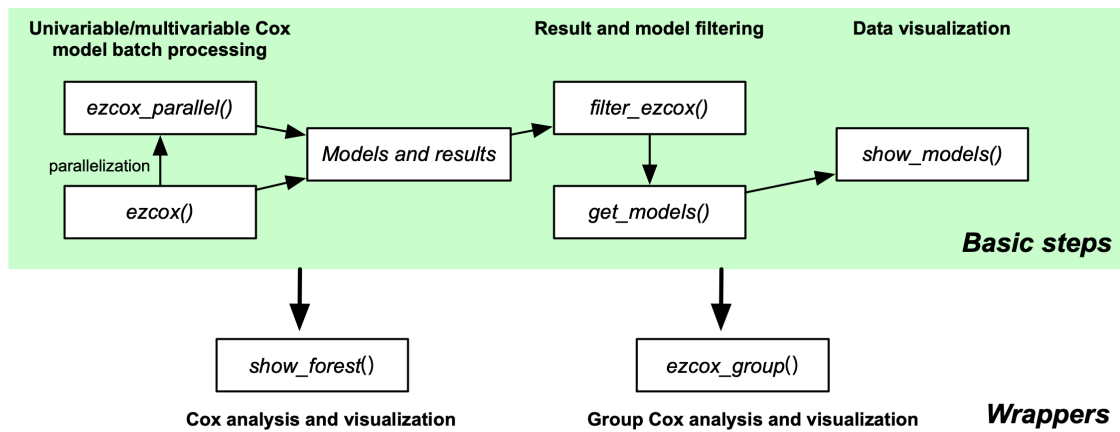

Fig. 1. The architecture diagram of *ezcox*.

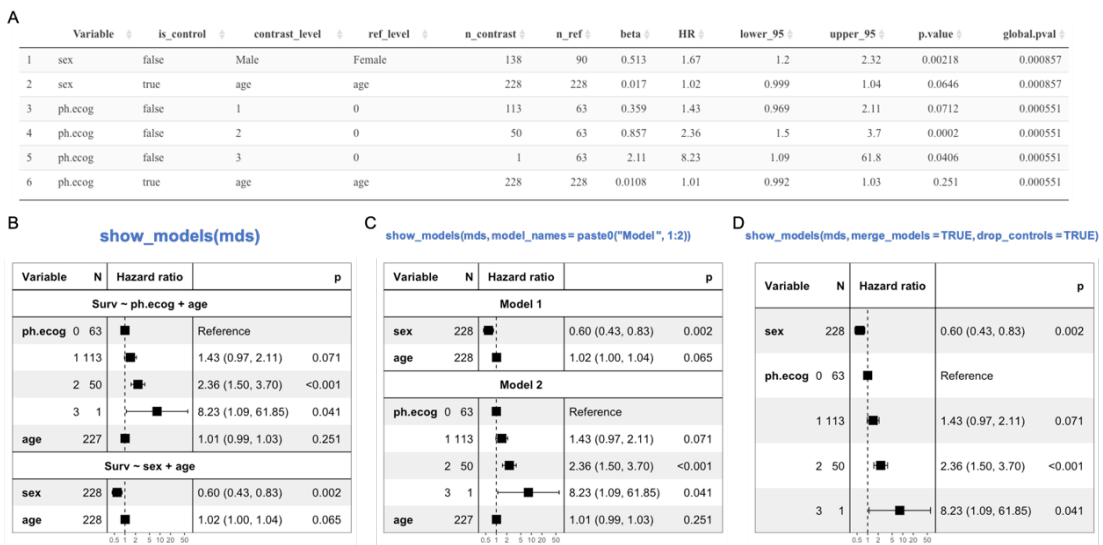

Fig. 2. Result data and forest plot generated by *ezcox*. Two multivariable Cox models are constructed, processed and visualized by *ezcox* with dataset "lung" from R package *survival* and command ezcox(lung, covariates = c("sex", "ph.ecog"), controls = "age", return_models=TRUE). (A) Tidy results from two models. Flexible forest plot to show the same model results with (B) default parameters, (C) modified model names and (D) selected variables (keep only the focal variables and drop control ones).

Based on the first part, we go further define two wrappers for quick data analysis. The first wrapper "*show_forest*" encapsulates functions in the first part to provide an end-to-end feature that users can generate final result plot from input data in just one command. In practical use, we found sometimes it is pretty common to explore one variable with Cox analysis in different groups. The second wrapper "*ezcox_group*" is designed to accomplish the task. It splits the data with groups and performs Cox analysis in each group. Data and models from all groups are combined and the result is also shown with forest plot (Fig. 3).

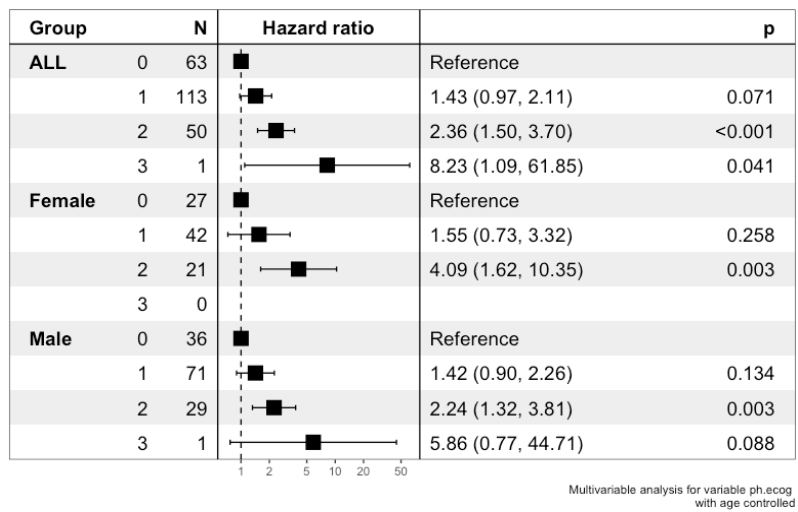

Fig. 3. A showcase of group Cox analysis result. The models are constructed, processed and visualized by *ezcox* with dataset "lung" from R package *survival* and command ezcox_group(lung, grp_var = "sex", covariate = "ph.ecog", controls = "age", add_all = TRUE). This forest plot shows how the variable "ph.ecog" affect patients' survival with age controlled in difference sex groups.

Molecular profile including mRNA expression, miRNA expression, transcript expression, protein expression, mutation status, methylation status, promoter activity, etc. are all suitable for *ezcox*. To illustrate how to use core features of *ezcox* for omics data, we provided a more practical vignette at https://shixiangwang.github.io/ezcox-adv-usage/ to show how to analyze mRNA expression of a gene list containing oncogenes and tumor suppressors in TCGA.

**Conclusion**

Here, we present *ezcox*, the first R package for Cox model batch processing and visualization. Since its release, *ezcox* has been downloaded for more than **14,000** times (according to the API for CRAN package download counts, from the RStudio CRAN mirror, https://cranlogs.r-pkg.org/) and utilized in multiple published cancer studies[10]–[13] around the world. We believe that *ezcox* could effectively promote the practical use of disease (e.g., cancer) related omics data and boost clinically relevant gene signature identification and biomarker discovery.

**Availability and implementation**

*ezcox* is an open source R package under GPL-3 license and its source code is hosted on https://github.com/ShixiangWang/ezcox. The stable release of ezcox is freely available at CRAN (https://cran.r-project.org/package=ezcox) and Conda forge channel (https://anaconda.org/conda-forge/r-ezcox). A ready web application hosted on Hiplot platform is available at https://hiplot.com.cn/basic/ezcox for users with little programming experience.

*Conflict of Interest*: none declared.